\documentclass[aps,pre,twocolumn,showpacs,groupedaddress]{revtex4}

\usepackage{graphicx}

 \newcommand {\dip}{\displaystyle}
 
 \newcommand {\f}{\dip \frac}
 \newcommand {\integ}{\displaystyle \int}
 
 \newcommand {\ig}{\includegraphics}

 \newcommand {\lp}{\mbox {\boldmath $($}}
 \newcommand {\rp}{\mbox {\boldmath $)$}}

\begin{document}


\title{Spectral analysis and  an area-preserving extension \\
 of a piecewise linear intermittent map}


\author{Tomoshige Miyaguchi}
\email{tomo@nse.es.hokudai.ac.jp}
\affiliation{%
Meme Media Laboratory, 
Hokkaido University, 
Kita-Ku, Sapporo 060-0813, Japan 
}
\author{Yoji Aizawa}
\affiliation{%
Department of Applied Physics, School of Science and
Engineering, Waseda University, 3-4-1 Okubo, Shinjuku-ku, Tokyo,
169-8555, Japan
} 

%


\date{\today}

\begin{abstract}
  We investigate spectral properties of a 1-dimensional piecewise linear
 intermittent map, which has not only a marginal fixed point but also a
 singular structure suppressing injections of the orbits into
 neighborhoods of the marginal fixed point. We explicitly derive
 generalized eigenvalues and eigenfunctions of the Frobenius--Perron
 operator of the map for classes of observables and piecewise constant
 initial densities, and it is found that the Frobenius--Perron operator
 has two simple real eigenvalues $1$ and $\lambda_d \in (-1,0)$, and a
 continuous spectrum on the real line $[0,1]$. From these spectral
 properties, we also found that this system exhibits power law decay of
 correlations. This analytical result is found to be in a good agreement
 with numerical simulations. Moreover, the system can be 
 extended to an area-preserving invertible map defined  on the unit
 square. This extended system is similar to the baker transformation,
 but does not satisfy hyperbolicity. A relation between this 
 area-preserving map and a billiard  system is also discussed.   
\end{abstract}


\maketitle


\section{Introduction}

In the  past decades, a lot of studies have been devoted to 
investigations of the relations between microscopic chaos and
non-equilibrium behaviors such as relaxation and transport, and it has
been found that microscopic chaos plays essential roles in
non-equilibrium processes \cite{gaspard2,dorfman}. 
For example, it is well known that, for fully chaotic (hyperbolic)
systems, correlation functions decay exponentially and their decay rates 
are characterized by the discrete eigenvalues of its Frobenius--Perron
(FP) operator \cite{ruelle1,ruelle2,ruelle3,pollicott2,pollicott4}. 
As one of the examples of the hyperbolic systems that permits detailed
calculations, the baker transformation has been studied extensively and
its spectral properties of FP operator are fully understood
\cite{hase1}. In addition, the baker map is considered as an
abstract model of chaotic Hamiltonian systems, because it has the
area-preserving property, which is an universal feature of the
Poincar\'e map of the Hamiltonian systems of 2 degrees of freedom
\cite{arnold1}. In fact, similarities of the baker map to the Lorentz
gas with finite horizon have been pointed out \cite{tel}.   

In contrast to hyperbolic systems, dynamics in generic Hamiltonian
systems is more complicated and diverse. When a phase space of a
Hamiltonian system consists of integrable (torus) and non-integrable
components (chaos), power law decay of correlations is frequently
observed  \cite{karney,chirikov,meiss1,meiss2,geisel3,zaslavsky}.
Although such kind of systems, i.e., systems with mixed type phase
spaces, are more generic than the integrable or the fully chaotic
systems, the theoretical understanding of their statistical properties
is not enough;~for example the ergodic and mixing properties of chaotic 
components of generic systems are still unclear from the theoretical
point of view.

For understanding sub-exponential decay of correlation functions in
dynamical systems, non-hyperbolic 1-dimensional maps have been studied by 
several authors 
\cite{geisel1,geisel2,artuso1,artuso2,hase3,tasaki1,tasaki2,schuster,aizawa,aaronson} 
and they have found power law decay of correlations in their models.
Therefore, it is natural to imagine connections of these non-hyperbolic
maps and mixed type Hamiltonian systems, however, extensions of these
maps to 2-dimensional area-preserving systems are unknown. 
Thus in this paper, we introduce a modified version of the 1-dimensional 
intermittent map studied in Ref.~\cite{tasaki1} and extend
it to an area-preserving system. This area-preserving map, which is
similar to the baker transformation, may be considered as an abstract
model of mixed type Hamiltonian systems. 

Our theoretical treatment is mainly based on Ref.~\cite{tasaki1}, where
a piecewise linear version of the Pomeau-Manneville map
\cite{pomeau1,pomeau2} is proposed and its generalized spectral
properties of the FP operator in a sense of
Refs.~{\cite{gelfand1,gelfand2}} have been elucidated.  Their model has 
a marginal fixed point and exhibits power law decay of correlation which
they have found to be an outcome of a continuous spectrum of the FP 
operator. In addition to a marginal fixed point, the piecewise linear
map studied in the present paper has a singular structure, which
suppresses injections of the orbits into neighborhoods of the marginal
fixed point. Due to this property, the uniform density is invariant
under time evolution and the map can be extended to an area-preserving  
map on the unit square. And it is shown that generalized eigenvalues of
the FP operator consists of two simple real eigenvalues $1$ and
$\lambda_d \in (-1,0)$, and a continuous spectrum on the real interval 
$[0,1]$. It is also shown that correlation functions exhibit power law 
decay due to the continuous spectrum.

This paper is organized as follows. In Sec.~\ref{sec:defs}, we introduce
the piecewise linear map, and define the FP operator and functional
spaces of observables and initial densities. In Sec.~\ref{sec:sp}, we
derive the spectral decomposition of the FP operator. In
Sec.~\ref{sec:extn}, long time behaviors are analyzed and some numerical   
results are displayed. We also discuss the extension of our model to an
area-preserving invertible map in Sec.~\ref{sec:extn}. 
Section~\ref{sec:summary} is devoted to summary and remarks that include
comments about differences between our model and the one in
Ref.~\cite{tasaki1}, and about similarities to a billiard system.      

 \begin{figure}
  \vspace*{.cm}
   \ig[width=8.6cm]{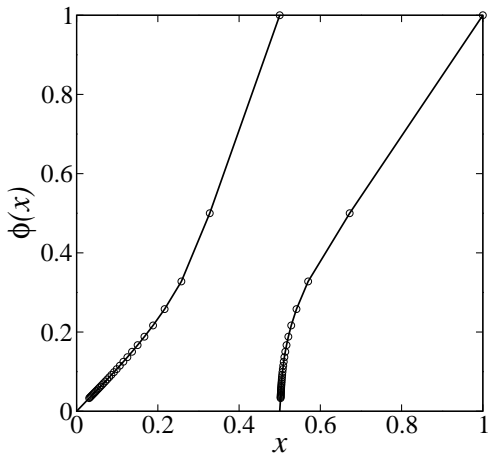} \vspace*{-.2cm}
   \caption{The piecewise linear map $\phi (x)$ (solid line) for $b=0.5$
   and $\beta =1.3$. The circles indicate endpoints of the straight
  line segments.}  
   \label{fig:cPMmap}
 \end{figure}

\section{\label{sec:defs}Piecewise linear map}
\subsection{Definition of the map}
 The dynamical system we consider in this paper is the piecewise linear 
 map shown in Fig.~\ref{fig:cPMmap}. This map $\phi (x):[0,1] \to [0,1]$
 consists of two parts:~the left $0 \leq x<b$ and the right part
 $b \leq x<1$. Each part is formed  by an infinite number of straight
 line segments;~the circles plotted in Fig.~\ref{fig:cPMmap}
 indicate endpoints of these segments. The map $\phi(x)$ is defined by  
 \begin{eqnarray}
  \phi(x) = 
   \left\{
    \begin{array}{l}
   \eta_{k}^{-} (x - \xi_{k}^{-}) + \xi_{k-1}^{-}  \\[.3cm]
    ~~~~~{\rm for~} x \in [\xi_{k}^{-},\xi_{k-1}^{-}) ~~~(k=1,2,\cdots),  \\[.5cm]
   \eta_{k}^{+} (x - \xi_{k}^{+}) + \xi_{k-1}^{-}   \\[.3cm]
    ~~~~~{\rm for~} x \in [\xi_{k}^{+},\xi_{k-1}^{+}) ~~~(k=1,2,\cdots).
   \end{array}
   \right.
   \label{eqn:1d-model}
 \end{eqnarray}
 In this definition, $\xi_k^-$~$(k=0,1,\cdots)$ represent the horizontal
 coordinates of the endpoints of the segments on the left  part ($x<b$)
 and they are defined as   
 \begin{eqnarray}
  \left\{
  \begin{array}{l}
   \xi_0^- = b, \\[.25cm]
   \xi_{k-1}^{-} - \xi_k^- = \f b{\zeta (\beta)} \left( \f 1k \right)^{\beta}
   ~~~{\rm for}~~k=1,2,\cdots,
   \end{array}
  \right.
 \end{eqnarray}
 where $\beta > 1$ is a parameter and 
 $\zeta (\beta) = \sum_{n=1}^{\infty} 1/n^{\beta}$ is the Riemann zeta
 function. And $\eta_k^-$ is the slope of the 
 $k$-th segment and  defined as 
 \begin{eqnarray}
  \eta_k^- &=& 
   \f {\xi_{k-2}^{-} - \xi_{k-1}^{-}}{\xi_{k-1}^{-} -  \xi_{k}^{-}} 
   \\[.25cm]
   &=&
   \left\{
    \begin{array}{ll}
      ~~\f {1-b}b \zeta (\beta)         &~~~{\rm for}~~k=1,  \\[.35cm]
      \left( \f k{k-1} \right)^{\beta}  &~~~{\rm for}~~k=2,3,\cdots,
    \end{array}
       \right.
 \end{eqnarray}
where we define $\xi_{-1}^- = 1$ for convenience.
This definition of the left part ($0<x<b$) is the same as that of the
piecewise linear Pomeau-Manneville map proposed in the 
Ref.~\cite{tasaki1}. The map $\phi(x)$ can be approximated as 
$\phi (x) \sim x + C x^{\beta/(\beta - 1)}$ when $x \to 0+$,
where $C$ is a constant. Thus the origin $x=0$ is a marginal fixed
point.  

In the same way, $\xi_k^+$~$(k=0,1,\cdots)$ are the horizontal
coordinates of the endpoints of the segments on the right part
and defined as 
\begin{eqnarray}
\left\{
\begin{array}{l}
\xi_1^+ = b\left( 1 + \f 1{\zeta (\beta)}\right), 
~~~~\xi_{0}^+ = 1,   \\[.5cm]
\xi_{k-1}^{+} - \xi_k^+ = \f b{\zeta (\beta)} 
\left( \f 1{(k-1)^{\beta}} - \f 1{k^{\beta}}  \right) \\[.35cm]
~~~{\rm for}~~k=2,3,\cdots,
\end{array}
\right.
\end{eqnarray}
and $\eta_k^+$ is the slope of the $k$-th segment and defined as
\begin{eqnarray}
\eta_k^+ &=& 
 \f {\xi_{k-2}^{-} - \xi_{k-1}^{-}}{\xi_{k-1}^{+} - \xi_{k}^{+}} \\[.25cm]
 &=&
\left\{
\begin{array}{lll}
\f {1-b}{1-b \left[1 + 1/{\zeta (\beta)}\right]}
&~~~{\rm for}~~k=1, \\[.35cm]
\f {k^{\beta}}{k^{\beta} - (k-1)^{\beta}} &~~~{\rm for}~~k=2,3,\cdots.
\end{array}
\right.
\end{eqnarray}
This is the definition of the right part of the map $\phi (x)$, 
$b \leq x < 1$. This part is different from the Pomeau-Manneville type maps,
and is similar to a map proposed by Artuso and Cristadoro \cite{artuso2}. 
$\phi(x)$ behaves as $\phi (x) \sim (x-b)^{(\beta-1)/{\beta}}$ when  
$x \to b+$. Therefore the derivative $\phi'(x)$ of the map is divergent
at $x=b+$. 

We assume that $\eta_1^+ > 0$, i.e., 
\begin{eqnarray}
b < \f {\zeta (\beta)}{ 1 + \zeta (\beta) }.
\label{eqn:condition}
\end{eqnarray}
Note that the uniform density on the interval $[0,1]$ is
invariant under the time evolution of this map because the relation   
$1/{\eta^+_k} + 1/{\eta^-_k}=1$ is satisfied for $k=1,2,\cdots$.  
Figure \ref{fig:cPMmap} shows the shape of the map $\phi(x)$ for
$\beta=1.3$ and $b = 0.5$. There is a singular structure near $x=b$, 
which suppresses injections of the orbits into neighborhoods of the
marginal fixed point $x=0$. This system can be easily extended to a
2-dimensional area-preserving map, whose dynamics of the expanding
direction is given by the map $\phi(x)$. This will be discussed in
Sec.~\ref{sec:extn}. 

\subsection{FP operator and functional spaces}
The FP operator $\hat P$ and its adjoint $\hat P^{\ast}$ 
are defined by 
\begin{eqnarray}
 \hat P \rho (x)   &=& 
  \integ_0^1 dy \delta  \lp x - \phi(y) \rp \rho (y), \\[.2cm]
  \hat P^{\ast} A(x) &=& A \lp \phi(x) \rp,
\end{eqnarray}
respectively. We also define an inner product $(A, \rho)$ as the average
of an observable $A(x)$ with respect to a density $\rho (x)$, 
\begin{equation} 
(A, \rho) = \integ_0^1 dx A(x) \rho(x).
\end{equation}
Then, the average of $A(x)$ at time $t$ with respect to an initial
density $\rho(x)$ is given by 
$(A, \hat P^t \rho) \equiv ( \hat P^{\ast t} A, \rho)$.

Let us consider an observable $A(x)$ such as the inequality 
\begin{equation}
|A(x) - a_0 - a_1 x| \leq K x^{\beta / (\beta - 1)} 
\label{eqn:assm-obs}
\end{equation}
holds for some positive constant $K$, where the constants $a_0$ and
$a_1$ satisfy $a_0=A(0)$ and $a_1=A'(0)$, respectively \cite{tasaki1}.  
This function is bounded on $[0,1]$ and smooth near the origin. And we
define a set $X_O$ as the functional space which consists of such
observables. This functional space is invariant under the action of the
adjoint of the FP operator $\hat P^{\ast}$, namely,  
$
\hat P^{\ast} A(x) \in X_O ~~{\rm if}~~A(x) \in X_O.
$
If we define the norm as
\begin{eqnarray} \nonumber
|| A(x) ||_O &=& |a_0| + |a_1| + \sup_x |A(x)| \\[.1cm]
&& + \sup_x  \f {|A(x) - a_0 - a_1 x|}{x^{\beta / (\beta - 1)}},
\end{eqnarray}
then this functional space becomes a Banach space with respect to this 
norm. Note that this space is dense in the Hilbert space $L^2[0,1]$ of
the square integrable functions on $[0,1]$. 

Furthermore we restrict initial densities to be piecewise constant
\cite{tasaki2},
\begin{equation}
 \rho (x) =    \tilde \rho_k  ~~~{\rm if}~~x \in [\xi^-_k, \xi^-_{k-1})
~~~(k=0,1,2,\cdots).
\end{equation}
We also assume the following properties for $k=1,2,\cdots$,
\begin{eqnarray}
\tilde \rho_k = \dip \sum_{l=0}^{\infty} \rho_l \left(\f k{k+l} \right)^{\beta}
~~~{\rm with}~~~
\dip \sum_{l=0}^{\infty} |\rho_l| \theta^l < + \infty,
\end{eqnarray}
where $\theta > 1$ is a constant. We also assume the normalization 
condition
\begin{eqnarray}
\f b{\zeta (\beta)} 
\dip \sum_{k=1}^{\infty} \f {\tilde \rho_k}{k^{\beta}}
+(1-b) \tilde \rho_0 = 1.
\label{eqn:normal}
\end{eqnarray}
In the following sections, we use this condition
[Eq.~(\ref{eqn:normal})] only for clarity of exposition.  But it is not 
essential and almost the same result can be obtained without it.  

We define a set $X_D$ as the functional space which consists
of such densities. This functional space is invariant under the action
of the FP operator;~namely,    
$
\hat P \rho \in X_D ~~~{\rm if}~~~\rho \in X_D.
$
And if the norm of this space is defined as
\begin{equation}
|| \rho ||_D = \sum_{l=0}^{\infty} |\rho_l| \theta^l,
\end{equation}
this functional space becomes a Banach space. The functional space $X_D$ 
is not dense in $L^2[0,1]$.

In the above definition for initial densities, we assume that the
initial densities are constant on the interval $[b,1)$, i.e., the right
part of the map. Although it seems to be a strong restriction and it is
possible to extend to the densities piecewise constant also in the right
part, this extension does not make any changes for the long time
behaviors because relaxation to a constant density in the right part
finishes by only one iteration of the map $\phi(x)$.

\section{\label{sec:sp}Spectral analysis}
The purpose of this section is to derive a spectral decomposition of 
the average $(A, \hat P^t \rho)$. First, we derive the matrix elements
of the resolvent operator of $\hat P$, then the average 
$(A, \hat P^t \rho)$ is obtained by an integral transformation of the  
matrix elements of the resolvent. Finally, deforming the integration
path, we have the spectral decomposition.

\subsection{Matrix elements of the resolvent operator}

Let us define the matrix elements of the resolvent of the
FP operator $\hat P$ as,
\begin{eqnarray} \nonumber
\left(
A, \f 1{z - \hat P} \rho
\right)
&=& \sum_{t=0}^{\infty} \f 1{z^{t+1}}
\left(
A, \hat P^t \rho
\right) \\[.cm]
&=& \sum_{k=1}^{\infty} \left[
\tilde \rho_k \hat B_k^- (z) + \tilde \rho_0 \hat B_k^+ (z)
\right],
\label{eqn:resolvent1}
\end{eqnarray}
where 
$\hat B_k^{\pm} (z)$ are defined below [see
Eq.~(\ref{eqn:hBk})]. Let us rewrite $(A, \hat P^t \rho)$ as  
\begin{eqnarray} \nonumber
\left(
A, \hat P^t \rho
\right)
&=& \integ_0^1 dx A \circ \phi^t (x) \rho(x) \\[.1cm]
&=& \dip \sum_{k=1}^{\infty} 
\left[
\tilde \rho_k B_k^- (t) + \tilde \rho_0 B_k^+ (t) 
\right],
\end{eqnarray}
where $B_k^{\pm}(t)$ are defined for $k=1,2,\cdots$ by
\begin{eqnarray}
B_k^{\pm} (t) \equiv 
\integ_{\xi_k^{\pm}}^{\xi_{k-1}^{\pm}} dx A \circ \phi^t (x).
\label{eqn:Bk}
\end{eqnarray}
Then $\hat B_k^{\pm}(z)$ are defined for $k=1,2,\cdots$ by 
\begin{eqnarray}
\hat B_k^{\pm}(z) \equiv \sum_{t=0}^{\infty} \f {B_k^{\pm} (t)}{z^{t+1}}.
\label{eqn:hBk}
\end{eqnarray}

From Eq.~(\ref{eqn:Bk}), we have the following recursion relations for
$B_k^{\pm}(t)$: 
\begin{eqnarray}
B_1^{\pm} (t+1) &=& \f 1{\eta_1^{\pm}} \sum_{k=1}^{\infty} B_{k}^+ (t), 
\label{eqn:recB1}\\[.1cm]
B_k^{\pm} (t+1) &=& \f 1{\eta_k^{\pm}} B_{k-1}^- (t)
~~~{\rm for}~~k=2,3,\cdots.
\label{eqn:recBk}
\end{eqnarray}

\noindent
From Eqs.~(\ref{eqn:hBk})--(\ref{eqn:recBk}), we
have the recursion relations for $\hat B_k^{\pm} (z)$:  
\begin{eqnarray}
\hat B_1^{\pm} (z) &=& \f {B_1^{\pm} (0)}z 
+ \f 1{z \eta_1^{\pm}} \sum_{k=1}^{\infty} \hat B_{k}^+ (z),
\label{eqn:hatB1pm} \\[.25cm]
\hat B_k^{\pm} (z) &=& \f {B_k^{\pm} (0)}z 
+ \f 1{z \eta_k^{\pm}} \hat B_{k-1}^- (z) 
~~~{\rm for}~~k=2,3,\cdots.
\nonumber\\[-.3cm]
\label{eqn:hatBkpm}
\end{eqnarray}
Using these relations recursively, the following equation can be derived
for $k=1,2,\cdots$: 
\begin{widetext}
\begin{eqnarray} 
\hat B_k^- (z) 
= \dip \sum_{m=0}^{k-1}
\f {B_{k-m}^- (0)}{z^{m+1}} \left( \f {k-m}{k} \right)^{\beta} +
\f 1{z^k k^{\beta}} \f {b}{(1-b) \zeta(\beta)}
\dip \sum_{m=1}^{\infty} \hat B_m^+ (z) .
\label{eqn:hatBk-}
\end{eqnarray}
Furthermore, from Eqs.~(\ref{eqn:hatBkpm}) and
(\ref{eqn:hatBk-}), we obtain for $k=2,3,\cdots$, 
\begin{eqnarray} 
\hat B_k^+ (z)
&=& \f {B_k^+ (0)}z + \f 1{\eta_k^+}
\left[ \dip \sum_{m=0}^{k-2} 
\f {B_{k-m-1}^- (0)}{z^{m+2}} \left( \f {k-m-1}{k-1} \right)^{\beta} +
\f 1{z^{k} (k-1)^{\beta}} \f {b}{(1-b) \zeta(\beta)}
\dip \sum_{m=1}^{\infty} \hat B_m^+ (z) 
\right] .
\label{eqn:hatBk+}
\end{eqnarray}
\end{widetext}
Summing up Eq.~(\ref{eqn:hatB1pm}) and Eqs.~(\ref{eqn:hatBk+}), we have
\begin{equation}
\dip \sum_{k=1}^{\infty} \hat B_k^+ (z) = \f{ \Phi (z)}{ Z(z)},
\label{eqn:PhiZ}
\end{equation}
where $Z(z)$ and $\Phi (z)$ are defined by
\begin{eqnarray}
Z (z) &\equiv& z - \f 1{\eta_1^+} - \f {b}{(1-b) \zeta(\beta)} 
 \dip \sum_{k=1}^{\infty} \f 1{\eta_{k+1}^+ z^{k} k^{\beta}} ,
\label{eqn:Z1} \\[.2cm]
\Phi (z) &\equiv& 
\dip \sum_{k=1}^{\infty} B_k^+ (0) 
+ \dip \sum_{l=1}^{\infty}  \dip \sum_{m=0}^{\infty} 
  \f {B_{l}^- (0)}{\eta_{m+l+1}^+ z^{m+1}} \left( \f {l}{m+l} \right)^{\beta}.
\nonumber
\\[-.2cm]
\label{eqn:Phi1} 
\end{eqnarray}
The functions $Z (z)$ and $\Phi (z)$ are absolutely convergent for
$|z| > 1$ and thus analytic there. From Eqs.~(\ref{eqn:hatBk-}),
and (\ref{eqn:PhiZ})--(\ref{eqn:Phi1}), the matrix elements of the
resolvent [Eq.~(\ref{eqn:resolvent1})] can be rewritten as 
\begin{eqnarray} \nonumber
\left(
A, \f 1{z - P} \rho
\right) &
=
\f {\Psi (z) \Phi (z)}{Z (z)}
+ \dip \sum_{l=1}^{\infty}  \dip \sum_{k=l}^{\infty} 
 \f {\tilde \rho_{k} B_{l}^- (0)}{z^{k-l+1}} \left( \f {l}k \right)^{\beta} 
\\[.15cm]
%
%
&\hspace*{-2.2cm} =
\f {\Psi (z) \Phi (z)}{Z (z)}
+  \f {(1-b) \zeta (\beta)}b 
\left[
\Xi (z) - B_{1}^- (0) \tilde \rho_0 
\right],
\label{eqn:resolvent2}
\end{eqnarray}
where we define the functions $\Psi (z)$ and $\Xi (z)$ as 
\begin{eqnarray}
\Psi (z) \equiv  \tilde \rho_0 + \f b{(1-b) \zeta (\beta)}
\sum_{k=1}^{\infty} \f {\tilde \rho_{k} }{z^{k} k^{\beta}},
\label{PSI1}
\end{eqnarray}
and
\begin{eqnarray} \nonumber
\Xi (z) &=& 
B_{1}^- (0) \Psi (z)  
+ \dip \sum_{l=2}^{\infty} B_{l}^- (0) l^{\beta} z^{l-1} \\[.cm]
&& \hspace*{-.3cm}
\times \left[
\Psi (z) - \tilde \rho_0 - \f b{(1-b) \zeta (\beta)} 
\dip \sum_{k=1}^{l-1} \f {\tilde \rho_{k} }{z^{k} k^{\beta}}
\right],
\label{XI1}
\end{eqnarray}
respectively. These functions $\Psi (z)$ and $\Xi(z)$ are also
absolutely convergent and analytic for $|z| > 1$. 

\subsection{Analytic properties of individual functions}
For $ r > 1$, we obtain the average $(A, \rho_t)$ 
by an integral transformation of the resolvent
[Eq.~(\ref{eqn:resolvent2})] as follows:
\begin{eqnarray} \nonumber
\left(
A,  \hat P^t \rho
\right)
&=& \dip \oint_{|z|=r} \f {dz}{2 \pi i} z^t
\left(
A, \f 1{z - P} \rho
\right)  \\[.25cm]
&&\hspace*{-0.9cm}= \dip \oint_{|z|=r} \f {dz}{2 \pi i} z^t
\left[
\f {\Psi (z) \Phi (z)}{Z (z)} + \f {(1-b) \zeta (\beta)}b \Xi (z)
\right], \nonumber
\\[-.12cm]
\label{eqn:resolvent3}
\end{eqnarray}
where the integration path is taken in the counter clockwise direction. 
In order to deform this integration path into the unit disk $|z|<1$ and
derive the spectral decomposition, we study the analytic properties of
the functions in the integrand of Eq.~(\ref{eqn:resolvent3}) in this
subsection. With the help of the identity 
\begin{eqnarray}
\f 1{k^{\beta}} = \f 1{\Gamma(\beta)} \integ_0^{\infty} ds~s^{\beta -1} e^{-ks},
\end{eqnarray}
we have analytic continuations of the functions $Z(z)$, $\Phi(z)$
and $\Psi(z)$ into the unit disk $|z|<1$. For example, $\Phi(z)$ can be
analytically continued as follows:
\begin{widetext}
\begin{eqnarray}
\Phi (z) &=& \dip \sum_{k=1}^{\infty} B_k^+ (0) 
+ \f 1{\Gamma (\beta)} \dip \sum_{l=1}^{\infty} 
  \f {B_{l}^- (0)  l^{\beta}}{z} 
  \integ_0^{\infty} ds s^{\beta -1} e^{-ls} (1-e^{-s}) 
  \dip \sum_{m=0}^{\infty} (z^{-1}e^{-s})^m
\\[.35cm]
&=& \dip \sum_{k=1}^{\infty} B_k^+ (0) +
\f 1{\Gamma (\beta)} \sum_{l=1}^{\infty} B_l^- (0) l^{\beta}
\integ_0^{\infty} ds \f {s^{\beta -1} e^{-ls}}{z - e^{-s}} (1-e^{-s}),
\label{eqn:Phi}
\end{eqnarray}
 where these calculations are justified because the convergence of the
 summation is uniform in $s$ for $|z| > 1$. Note that each integral in
 the right-hand side (rhs) of Eq.~(\ref{eqn:Phi}) is analytic except for
 the real interval $[0,1]$. Then the function $\Phi(z)$ expressed as
 Eq.~(\ref{eqn:Phi}) is analytic except for the cut $[0,1]$, because the
 second infinite sum in Eq.~(\ref{eqn:Phi}) is absolutely convergent, i.e.,
\begin{eqnarray} 
\f 1{\Gamma (\beta)} \sum_{l=1}^{\infty}
\left| B_l^- (0) l^{\beta}
\integ_0^{\infty} ds \f {s^{\beta -1} e^{-ls}}{z - e^{-s}} (1-e^{-s})
\right| 
\leq \f {b \dip \sup_x |A(x)| }{\zeta (\beta) d(z,[0,1])},
\end{eqnarray}
where $d(z, [0,1])$ is the distance between $z$ and the real interval 
$[0,1]$. 

In the same way, we obtain analytical continuations of the  functions
 $Z(z)$ and $\Psi(z)$:
\begin{eqnarray}
Z(z) &=& (z-1) 
\left[
1 + \f b{(1-b) \zeta (\beta) \Gamma (\beta)}
\integ_0^{\infty} ds \f {s^{\beta -1} e^{-s}}{z - e^{-s}}
\right], \\[.3cm]
\Psi (z) &=& \tilde \rho_0 + \f b{(1-b)\zeta (\beta) \Gamma(\beta)} 
\dip \sum_{l=0}^{\infty} \rho_l
\integ_0^{\infty} ds \f {s^{\beta -1} e^{-(l+1)s}}{z - e^{-s}}.
\end{eqnarray}
These expressions for the functions $Z(z)$ and $\Psi (z)$ are also
 analytic except for the cut $[0,1]$.
And the function $\Xi (z)$ can be also analytically continued into the
unit disk $|z|<1$ in the same way. Here, however, we rewrite the
 function $\Xi(z)$ in the following form: 
\begin{eqnarray}
\Xi (z) = \Psi(z) \dip \sum_{l=1}^{\infty} l^{\beta} B_l^- (0) z^{l-1} 
 -  \dip \sum_{l=2}^{\infty} l^{\beta} B_l^- (0) z^{l-1}
\left[ 
\tilde \rho_0 + \f b{(1-b)\zeta (\beta)} 
\sum_{k=1}^{l-1} \f {\tilde \rho_k}{z^k k^{\beta}}
\right]. 
\end{eqnarray}
The infinite sums in this expression of the function $\Xi (z)$ are 
absolutely convergent for $|z| < 1$. \vspace*{.3cm}
\end{widetext}


Let us consider the zeros of the function $Z (z)$. First we 
define a function $\Omega (z)$ as $\Omega (z) \equiv {Z(z)}/(z-1)$.
And if ${\rm Im} (z) \ne 0$, then
\begin{equation}
{\rm Im~}\Omega (z)
= - \f {b {\rm Im(z)}}{(1-b) \zeta (\beta) \Gamma (\beta)}
\integ_0^{\infty} ds \f {s^{\beta-1} e^{-s}}{|z - e^{-s}|^2} \ne 0.
\end{equation}
And also if ${\rm Im} z = 0$ and ${\rm Re} z > 1$, then $\Omega(z)>0$,  
~because $z-e^{-s} > 0$. Thus there is not zero of $Z(z)$ in these
regions. 

Next if ${\rm Im}z = 0$ and ${\rm Re} z < 0$, we have
\begin{eqnarray}
\Omega'(z) = 
- \f b{(1-b) \zeta (\beta) \Gamma (\beta)}
\integ_0^{\infty} ds \f {s^{\beta-1} e^{-s}}{(z - e^{-s})^2} < 0,
\end{eqnarray}
and
\begin{eqnarray}
\begin{array}{lll}
\Omega(-1) &=& 1 - \f b{(1-b) \zeta (\beta) \Gamma (\beta)}
\integ_0^{\infty} ds \f {s^{\beta-1} }{e^s + 1} \\[.5cm] 
&>& 1 - \f b{(1-b) \zeta (\beta) \Gamma (\beta)} 
\integ_0^{\infty} ds \f {s^{\beta-1} }{e^s}   \\[.5cm]
&=& 1 - \f b{(1-b) \zeta(\beta)} > 0
\end{array}
\end{eqnarray}
The last inequality holds because of the
Eq.~(\ref{eqn:condition}). And it is easy to see that 
$\Omega (z) \to - \infty$ as $z \to 0$.
Therefore, on the real interval $[-1,0]$, $\Omega(z)$ has the unique
zero $\lambda_d$ of order 1.

\subsection{Spectral decomposition}

From the results of the last subsection, the contour of the integration of
Eq.~(\ref{eqn:resolvent3}) can be deformed into the unit disk $|z| < 1$ as
\begin{eqnarray} \nonumber
\left(
A,  \hat P^t \rho
\right)
&=& \dip \lim_{\eta \to 0} \dip \oint_{|z-\lambda_d|= \eta} \f {dz}{2 \pi i} z^t
\f {\Psi (z) \Phi (z)}{Z (z)} 
\hspace*{2.cm} \\[.38cm]
&&
\hspace*{-1.5cm}
+ \dip \lim_{\eta \to 0} \dip \oint_C  \f {dz}{2 \pi i} z^t
\left[
\f {\Psi (z) \Phi (z)}{Z (z)} + \f {(1-b) \zeta (\beta)}b \Xi (z)
\right],
\label{eqn:average1}
\end{eqnarray}
where the integration path $C$ is defined by
\begin{eqnarray} \nonumber
C \equiv 
\left\{
~z~ |~ |z| = \eta,~|z-1|=\eta~ 
\right. \hspace*{2cm} \\[.28cm] 
\left.
{\rm or}~z=\lambda \pm i0  ~(\eta < \lambda < 1 - \eta)
\right\}.
\end{eqnarray}
The first term of the rhs of Eq.~(\ref{eqn:average1}) corresponds to the 
simple pole at $z= \lambda_d$ and thus can be calculated by the residue
theorem and Eq.~(\ref{eqn:average1}) is rewritten as 
\begin{eqnarray}\nonumber
\left(
A,  \hat P^t \rho
\right)
&=& {\lambda_d}^t
\f {\Psi (\lambda_d) \Phi (\lambda_d)}{(\lambda_d - 1) \Omega'
(\lambda_d)} 
\hspace{3.7cm} \\[0.38cm]
&& \hspace{-1.5cm}
+ \dip \lim_{\eta \to 0} \dip \oint_C  \f {dz}{2 \pi i} z^t
\left[
\f {\Psi (z) \Phi (z)}{Z (z)} + \f {(1-b) \zeta (\beta)}b \Xi (z)
\right]. 
\label{eqn:average2}
\end{eqnarray}

Let us consider the integral of rhs of Eq.~(\ref{eqn:average2}).
Because of the facts (see the appendix)  
\begin{eqnarray}
\dip \lim_{z \to 0} & z^{\alpha} \Xi (z) = 0,  
\label{eqn:xito0}
\end{eqnarray}
and
\begin{eqnarray}
\dip \lim_{z \to 0} & z^{\alpha} \f {\Phi (z) \Psi (z)}{Z (z)} = 0,
\label{eqn:zto0}
\end{eqnarray}
for ${^{\forall}}\alpha > 0$, the contribution from the integration
around the origin, $\{ |z|=\eta \}$, vanishes. 

Similarly, we obtain (see the appendix) 
\begin{eqnarray}
\lim_{z \to 1} (z-1) \Xi (z) = 0,
\label{eqn:xito1}
\end{eqnarray}
and then the contribution from the integral around $z=1$,
$\{|z-1|=\eta \}$, vanishes for the second term of the integrand. On the
other hand, we have 
\begin{eqnarray} \nonumber
\dip \lim_{\eta \to 1} \integ_{|z-1|=\eta} dz 
 z^t \f {\Psi (z) \Phi (z)}{Z (z)} 
\\[0.3cm] \nonumber
&&\hspace*{-3cm}=
 \dip \lim_{z \to 1}  (z-1) \f {\Phi (z) \Psi (z)}{Z (z)}  
\\[0.3cm] \nonumber
&&\hspace*{-3cm}=  \integ_0^{1} dx A(x) 
\left[
 (1-b) \tilde \rho_0 + \f {b}{\zeta (\beta)}
\sum_{k=1}^{\infty} \f {\tilde \rho_k}{k^{\beta}}
\right]
\\[.3cm]
&&\hspace*{-3cm}=  \integ_0^{1} dx A(x), \\[-.05cm] \nonumber
\label{eqn:equilibrium}
\end{eqnarray}
where we have used the normalization condition Eq.~(\ref{eqn:normal}). 
The rhs of Eq.~(\ref{eqn:equilibrium}) is the average of $A(x)$ with
respect to the invariant density which is uniform for the map
$\phi(x)$;~therefore Eq.~(\ref{eqn:equilibrium}) gives the average value
of the invariant state. 

Finally we have to evaluate the integral along the cut. For this
purpose, we define the new functions 
$\hat \Omega (\lambda), ~f_{\lambda}(x)$ 
and $\nu_l(x)$ for $0< \lambda < 1$ as follows,
\begin{eqnarray}
\hat \Omega (\lambda) &\equiv& 
1 + \f {b}{(1-b) \zeta (\beta) \Gamma (\beta)}
\integ_0^{\infty} \hspace*{-.2cm} 
ds {\mathcal P} \f {s^{\beta -1} e^{-s}}{\lambda -e^{-s}} ,
\hspace*{1.cm}\\[.5cm]
\nonumber
f_{\lambda} (x) &\equiv& 
\f 1{1-b} 
\left[
\chi_0 (x)  + 
\f 1{\Gamma (\beta)} \sum_{l=1}^{\infty} l^{\beta} \chi_l (x) 
\right. 
\\[.3cm]
&&
\left.
 \hspace*{1.3cm} \times \integ_0^{\infty} ds {\mathcal P} 
 \f {s^{\beta -1} e^{-ls}}{\lambda - e^{-s}} (1-e^{-s}) 
\right] , \\[.5cm]
\nu_l (\lambda) &\equiv&
\f 1{\zeta (\beta) \Gamma (\beta)} 
\integ_0^{\infty} ds {\mathcal P} \f {s^{\beta -1} e^{-(l+1)s}}{\lambda- e^{-s}} ,
\end{eqnarray}
where ${\mathcal P}$ means the Cauchy's principle value and $\chi_l(x)$ is
defined, for $l=0,1,\cdots$, by 
\begin{eqnarray}
\chi_l(x) = 
\left\{
\begin{array}{ll}
1 & ~~~{\rm for}~~\xi^-_l \leq x < \xi^-_{l-1}, \\[.5cm]
0 & ~~~{\rm otherwise}.
\end{array}
\right.
\end{eqnarray}
Note that we have defined $\xi_{-1}=1$ in Sec.~{\ref{sec:defs}}. These
functions are related to the real parts of the functions  
$\{Z(z), \Phi(z), \Psi(z)\}$ near the cut:~
${\rm Re} Z(\lambda \pm i0) = (\lambda -1) \hat \Omega (\lambda)$, 
${\rm Re} \Phi(\lambda \pm i0) = (1-b)\int_0^1 dx A(x) f_{\lambda}(x)$, and
${\rm Re} \Psi(\lambda \pm i0) = 
 \tilde \rho_0 + b/(1-b) \sum_{l=0}^{\infty} \rho_l \nu_l(\lambda)$.

It can also be shown for the imaginary parts of the functions  
$\{Z(z), \Phi(z), \Psi(z)\}$ near the cut that
\begin{eqnarray}
\f {{\rm Im} \Phi (\lambda - i0)}{{\rm Im} Z (\lambda + i0)}
&=&  \f {(1-b) \zeta (\beta)}b 
    \sum_{l=1}^{\infty} l^{\beta} B_l^-(0) \lambda^{l-1}, \\[.5cm]
\f {{\rm Im} \Psi (\lambda - i0)}{{\rm Im} Z (\lambda + i0)}
&=&  \f 1{1-\lambda}
    \sum_{l=0}^{\infty} \rho_l \lambda^l.
\end{eqnarray}

Using these functions, we can calculate the integral along the cut as
follows \cite{tasaki1}:   
\begin{widetext}
\begin{eqnarray}\nonumber
\dip \lim_{\eta \to 0} && \hspace*{-.3cm} 
\integ_{C \backslash \{|z|=\eta ~\cup ~|z-1|=\eta \}}
\f {dz}{2 \pi i} z^t 
\left[
\f {\Psi (z) \Phi (z)}{Z (z)} + \f {(1-b)\zeta (\beta)}b \Psi (z)
\sum_{l=1}^{\infty} l^{\beta} B_l^- (0) z^{l-1}
\right]  \\[0.3cm]
&=& \integ_0^1 \f {d \lambda}{\pi} \lambda^t 
\f {{\rm Im} \left[ Z (\lambda + i0) \Phi (\lambda - i0) \right] 
    {\rm Im} \left[ Z (\lambda + i0) \Psi (\lambda - i0) \right] }
   { |Z (\lambda + i0)|^2 {\rm Im}~Z(\lambda + i0) }
\\[.3cm]
&=& \f b{(1-b) \zeta (\beta) \Gamma (\beta)} 
\integ_0^1 d \lambda 
\f { {\rm Im} \left[ Z (\lambda + i0) \Phi (\lambda - i0) \right] }
   { {\rm Im}~Z(\lambda + i0) } ~ 
\f { \lambda^t (1 - \lambda) \left( \log \frac 1{\lambda} \right)^{\beta - 1} }
   { |Z (\lambda + i0)|^2 }~
\f { {\rm Im} \left[ Z (\lambda + i0) \Psi (\lambda - i0) \right] }
   { {\rm Im}~Z(\lambda + i0) } ~  \\[.3cm]
&=& \integ_0^1 d \lambda ~(A, F_{\lambda}) \lambda^t (\tilde F_{\lambda}, \rho),
\label{eqn:cut}
\end{eqnarray}
where we define the linear functional $(A, F_{\lambda})$ of the observables
$A(x)$ as 
\begin{eqnarray} 
(A, F_{\lambda}) \equiv
N (\lambda) \integ_0^1 dx \{ A(x) - A(0) \} 
\left[
 f_{\lambda} (x)  
 + \f {(\lambda - 1) \hat \Omega (\lambda) \zeta (\beta)}b
 \dip \sum_{l=1}^{\infty} l^{\beta} {\lambda}^{l-1} \chi_l (x)
\right].
\label{eqn:l-func-ob}
\end{eqnarray}
The function $N(\lambda)$ in Eq.~(\ref{eqn:l-func-ob}) is given by 
\begin{eqnarray}
N(\lambda) \equiv
 \frac { \frac b{\zeta (\beta) \Gamma (\beta) (1-b)}   
\left( \log \frac 1{\lambda} \right)^{\beta - 1} }
{
(1 - \lambda) \left\{
\hat \Omega^2 (\lambda) + 
 \left[
\frac {b}{\zeta (\beta) \Gamma (\beta) (1-b)}
\left( \log  \frac 1 {\lambda} \right)^{\beta -1}  
\right]^2 \right\} }
\end{eqnarray}
\end{widetext}
We also define in Eq.~(\ref{eqn:cut}) the linear functional 
$(\tilde F_{\lambda}, \rho)$ of the initial densities $\rho (x)$ as   
\begin{equation}
( \tilde F_{\lambda}, \rho ) \equiv
(1-b) \tilde \rho_0 + b \dip \sum_{l=0}^{\infty} \rho_l
\left[
\nu_l (\lambda) - \f {1-b}b \hat \Omega (\lambda) \lambda^l
\right].
\end{equation}

Similarly, the linear functionals associated to the eigenvalues 
$z=\lambda_d$ and $z=1$, that have been obtained in 
Eqs.~(\ref{eqn:average2}) and (\ref{eqn:equilibrium}), can be
expressed as 
\begin{eqnarray}
(A, F_{d}) &\equiv& \f {1}{(\lambda_d - 1) \Omega'(\lambda_d)} 
\integ_0^1 dx A(x) f_{\lambda_d}(x),               \\[.1cm]
( \tilde F_{d}, \rho ) &\equiv& 
(1-b) \tilde \rho_0 + 
b \dip \sum_{l=0}^{\infty} \rho_l \nu_l (\lambda_d), 
\end{eqnarray}
and
\begin{eqnarray}
(A, F_{\rm in})    &\equiv& \integ_0^1 dx A(x),    \\[.1cm]
( \tilde F_{\rm in}, \rho ) &\equiv& 1,
\end{eqnarray}
respectively. It is easy to check that $(A, F_{\lambda}) = (A, F_{d}) = 0$~ if
$A(x)$ is a constant function and that
$(\tilde F_{\lambda}, \rho) = (\tilde F_{d}, \rho) = 0$ 
if $\rho (x) \equiv 1$.

Using these linear functionals, we can spectrally decompose 
the average $(A, \rho_t)$ [Eq.~(\ref{eqn:average2})] at time 
$t = 0, 1, 2, \cdots$ as,  
\begin{eqnarray} \nonumber
(A,\hat P^t \rho) &=& (A, F_{\rm in})(\tilde F_{\rm in}, \rho)  
+ {\lambda_d}^t (A, F_{d})(\tilde F_{d}, \rho)  \hspace*{1cm} \\[.25cm]
&&+ \integ_0^1 d \lambda \lambda^t
(A, F_{\lambda})(\tilde F_{\lambda}, \rho).
\label{eqn:sp-dc}
\end{eqnarray}
The left eigenfunctions 
$\{ F_{\rm in}, F_{\rm d}, F_{\lambda} \}$ 
satisfy the relations
\begin{eqnarray}
\begin{array}{lllll}
(A,\hat P F_{\rm in}) 
&\equiv& (\hat P^{\ast} A, F_{\rm in})
&=&      (A, F_{\rm in}), \\[.25cm]
(A,\hat P F_{\rm d}) 
&\equiv& (\hat P^{\ast} A, F_{\rm d})
&=&      \lambda_d(A, F_{\rm d}), \\[.25cm]
(A,\hat P F_{\lambda}) 
&\equiv& (\hat P^{\ast} A, F_{\lambda}) 
&=&      \lambda (A, F_{\rm \lambda}).
\end{array}
\end{eqnarray}
Therefore $\{ F_{\rm in}, F_{\rm d}, F_{\lambda} \}$ are 
eigenfunctions of the FP operator in a generalized sense
\cite{gelfand1,gelfand2,tasaki1,tasaki2}.
On the other hand, the right eigenfunctions 
$\{ \tilde F_{\rm in}, \tilde F_{\rm d}, \tilde F_{\lambda} \}$ 
satisfy the relations
\begin{eqnarray}
\begin{array}{lllll}
(\hat P^{\ast} \tilde F_{\rm in}, \rho) 
&\equiv& (\tilde F_{\rm in}, \hat P \rho) 
&=&      (\tilde F_{\rm in }, \rho), \\[.25cm]
(\hat P^{\ast} \tilde F_{\rm d},  \rho) 
&\equiv& (\tilde F_{\rm d}, \hat P \rho) 
&=& \lambda_d (\tilde F_{\rm d}, \rho), \\[.25cm]
(\hat P^{\ast} \tilde F_{\lambda}, \rho) 
&\equiv& (\tilde F_{\lambda}, \hat P \rho) 
&=& \lambda (\tilde F_{\lambda}, \rho) .
\end{array}
\end{eqnarray}
Therefore 
$\{ \tilde F_{\rm in}, \tilde F_{\rm d}, \tilde F_{\lambda} \}$  
are eigenfunctions of the adjoint of the FP operator in a generalized
sense. 

\begin{figure*}
 \vspace*{-.cm}
   \ig[width=15.7cm,height=7.5cm]{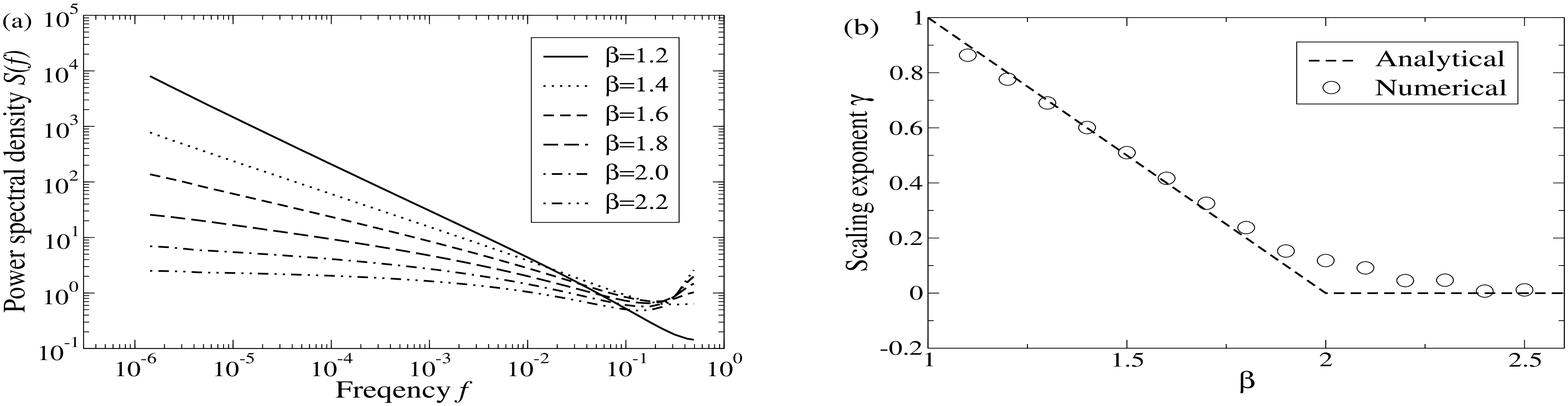} \vspace*{-.1cm}
 \caption{(a)~Power spectral densities $S(f)$ of time series $A(x(t))$
 (in log-log form) for five different values of the system parameter
 $\beta: \beta = $1.2~(the solid line), 1.4~(dotted), 1.6~(broken),
 1.8~(dashed), 2.0~(dashed-dotted), 2.2~(dashed-two-dotted). These PSDs
 are obtained by averaging over 20000 initial conditions uniformly
 distributed in $[0,1]$. 
 ~(b) The scaling exponent $\gamma$ of the PSD 
 $S(f)  \sim 1/f^{\gamma}$ as a function of $\beta$. The
 circles are the numerical results obtained by least square 
 fitting in the low frequency region (below $f=10^{-4}$) of the PSDs
 $S(f)$;~and the dashed line is the theoretical prediction
 [Eq.~(\ref{eqn:psd})]. }
 \vspace{.cm}
\label{fig:PSD-NHBmap}
\end{figure*}

Thus we obtain the spectral decomposition Eq.~(\ref{eqn:sp-dc}), where 
the spectrum consists of two discrete eigenvalues $1$ and $\lambda_d$,
and a continuous spectrum on $[0,1]$. And Eq.~(\ref{eqn:sp-dc}) for 
$t = 0$ shows that these eigenfunctions are complete.

\section{\label{sec:extn} 
Long time behaviors and an area-preserving extension} 

\subsection{Long time behaviors}

In the previous section, we derive the spectral decomposition of the
average $(A, \hat P^t \rho)$ and found that there is a continuous
spectrum. Since long time behaviors are controlled by eigenvalues whose  
absolute values are close to 1, we consider the limit $\lambda \to 1$
for the continuous spectrum. The eigenstate associated with the
eigenvalue $\lambda_d$ does not contribute to long time behaviors,
because this state decays exponentially. Under the assumption
Eq.~(\ref{eqn:assm-obs}), the leading term of the left eigenstate $(A,
F_{\lambda})$ when  $\lambda \simeq 1$ is given by 
\begin{equation}
(A, F_{\lambda}) \simeq K (1- \lambda)^{\beta - 2}
\integ_0^1 dx  [ A(x) -A(0) ]
\label{eqn:cut-asymp1}
\end{equation}
where $K$ is a constant. Similarly, we obtain the leading term for the
right eigenstate $(\tilde F_{\lambda}, \rho)$,
\begin{eqnarray} \nonumber
(\tilde F_{\lambda}, \rho) \simeq
(\tilde F_{\rm in}, \rho) - \sum_{l=0}^{\infty} \rho_l 
= 1 - \sum_{l=0}^{\infty} \rho_l
\label{eqn:cut-asymp2}
\end{eqnarray}
as $\lambda \simeq 1$. Using these facts and Eq.~(\ref{eqn:sp-dc}), we
have for $t \to \infty$ 
\begin{eqnarray} \nonumber
(A, \hat P^t \rho) &\simeq&
\integ_0^1 dx A(x)  \\[.2cm]
 && \hspace*{-1.2cm}+ \f {K'}{t^{\beta - 1}} 
\left( 1 - \dip \sum_{l=0}^{\infty} \rho_l \right)
\integ_0^1 dx  [ A(x) - A(0) ],
\label{eqn:correlation}
\end{eqnarray}
where $K'$ is a constant. Eq.~(\ref{eqn:correlation}) shows that the
correlation functions decay algebraically.
From Eq.~(\ref{eqn:correlation}), it is found that the power spectral
density (PSD) $S(f)$ behaves as \cite{schuster}
\begin{eqnarray}
S(f) \sim 
\left\{
\begin{array}{lll}
\f 1{f^{2-\beta}} &~~{\rm for}~~& 1 < \beta < 2, \\[.39cm]
|\log f|          &~~{\rm for}~~& \beta = 2,     \\[.39cm]
{\rm const.}      &~~{\rm for}~~& \beta > 2. 
\label{eqn:psd}
\end{array}
\right.
\end{eqnarray}

Figure~\ref{fig:PSD-NHBmap}(a) shows PSDs $S(f)$ of time series
$A(x(t))$, where the observable $A(x)$ is a step function defined on 
$x \in [0,1)$ as  
\begin{eqnarray}
A(x) = 
\left\{
\begin{array}{lll}
-1 &~~{\rm for}~~& x \in [0, 1/2) \\[.25cm]
 1 &~~{\rm for}~~& x \in [1/2, 1).
\end{array}
\right.
\end{eqnarray}
And $x(t)$ is produced by the successive iterations of the map $\phi(x)$:
$x(t+1) = \phi (x(t))$. Each PSD is obtained by averaging over 20000 
 initial conditions uniformly distributed in $[0,1]$. 


\begin{figure*}
 \ig[width=15.0cm,height=8.8cm]{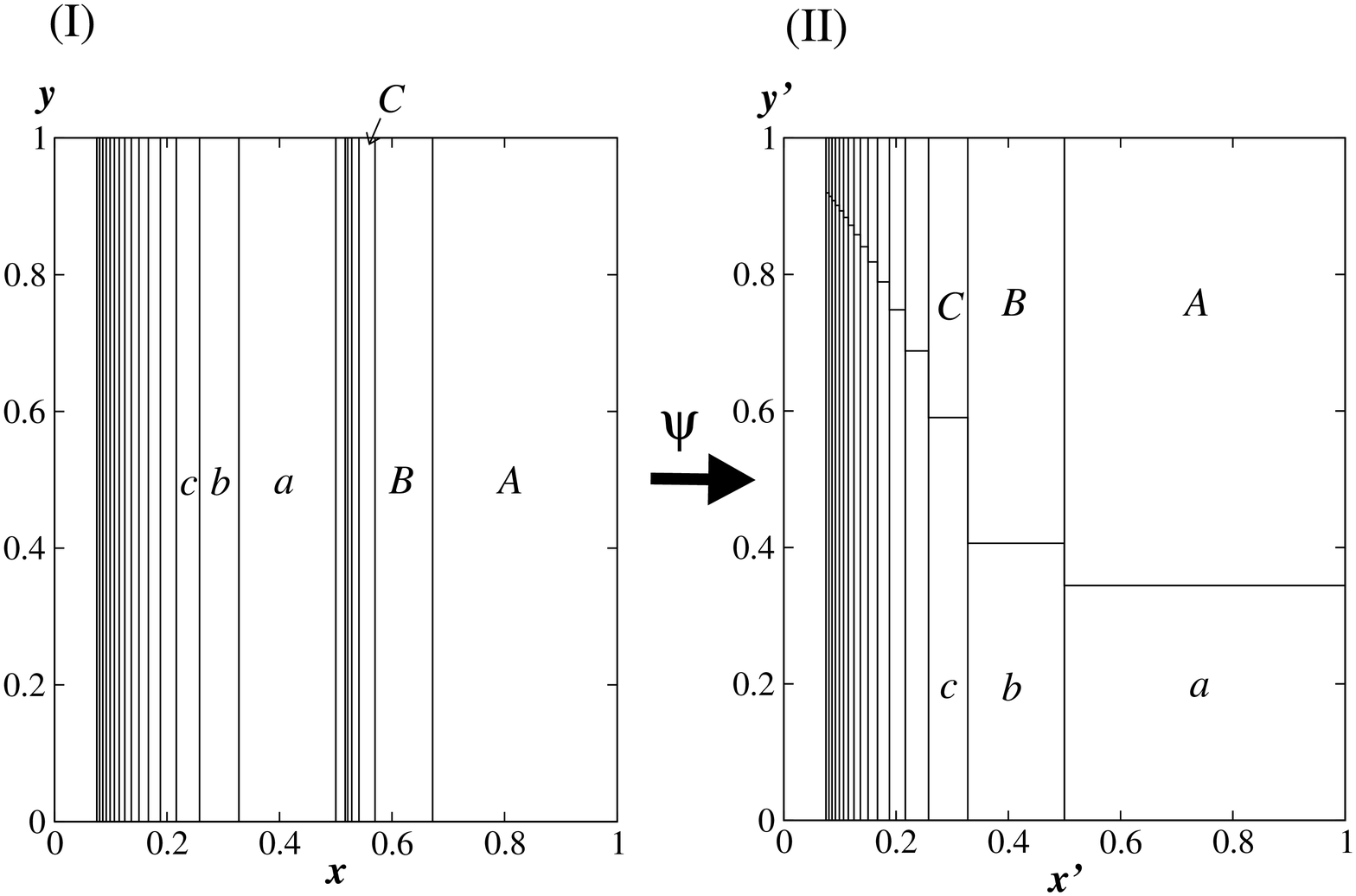} \vspace*{-.cm}
 \caption{ The area-preserving map $\psi(x,y)$ defined by
 Eq.~(\ref{eqn:2d-model}) for $b=0.5$ and $\beta=1.3$. The domains with
 labels  \{A, B, C, a, b, c\} in the left cell (I) are mapped into  the 
 domains with the same labels in the right (II), respectively. 
 Each domain is uniformly stretched in the horizontal direction and
 uniformly squeezed in the vertical direction. There are infinite number
 of such domains and each one is mapped in a similar way.  The partition
 is displayed only for the region $\{x \in [\xi_{15}^-,\xi_{0}^-]\}$ and 
 $\{ x \in  [\xi_{6}^+,\xi_{0}^+]\}$ in the left cell (I); only for 
 $x \in  [\xi_{15}^-, \xi_0^+]$ in the right (II).  The other regions
 are not displayed because the structures are too fine to see. }
\label{fig:NHBmap}
 \vspace{-.cm}
\end{figure*}

 In Fig.~\ref{fig:PSD-NHBmap}(a), the PSDs for $\beta < 2$ exhibit clear
 $1/f^{\gamma}$ scalings in low frequency regions.  Figure
 \ref{fig:PSD-NHBmap} displays this scaling exponent $\gamma$ of the PSD 
 $S(f) \sim 1/f^{\gamma}$ as a function of the system parameter
 $\beta$. And we show the theoretical prediction derived in the above 
 [Eq.~(\ref{eqn:psd})] by the dashed line in
 Fig.~\ref{fig:PSD-NHBmap}(b). Obviously, the numerical results show  a
 good agreement with the theoretical prediction.     

\subsection{Area-preserving extension}

As stated in the introduction, the 1 dimensional map investigated in the
present paper can be extended to an area-preserving 2-dimensional
transformation defined on the unit square; this extension 
$\psi: [0,1)^2 \to [0,1)^2$ is defined as
\begin{eqnarray}
\psi (x, y) = 
\left\{
\begin{array}{l}
\left(
\eta_{k}^{-} (x - \xi_{k}^{-}) + \xi_{k-1}^{-},~
\f y{\eta_k^-}
\right)  \\[.38cm]
\hspace*{2.7cm}  
{\rm for} ~x \in [\xi_{k}^{-},\xi_{k-1}^{-}),  \\[.33cm]
\left(
 \eta_{k}^{+} (x - \xi_{k}^{+}) + \xi_{k-1}^{-},~
 \f y{\eta_k^+} + \f 1{\eta_k^-}
\right)   \\[.38cm]
\hspace*{2.7cm}
{\rm for} ~x \in [\xi_{k}^{+},\xi_{k-1}^{+}).
\end{array}
\right.
\label{eqn:2d-model}
\end{eqnarray}
where $k=1,2,\cdots$. Obviously, the transformation for the horizontal 
coordinate $x$ is the same as the map $\phi(x)$ defined by
Eq.~(\ref{eqn:1d-model}) and does not depend on the vertical coordinate
$y$. This relation between 1-dimensional map $\phi(x)$ and its
area-preserving extension $\psi(x,y)$ is the same as that between the 
Bernoulli and the baker transformations. Note that the map $\psi(x,y)$
is area-preserving, because the Jacobian of this map equals to $1$
everywhere. The phase space (i.e., the unit square) of this
area-preserving map can be partitioned into infinite pieces like 
\begin{eqnarray} \nonumber
[0,1)^2 &=& 
\bigcup_{k=1}^{ \infty} \{(x,y) ~|~ x \in [\xi_k^-, \xi_{k-1}^-), y \in [0,1) \} 
\hspace*{1cm} \\[.0cm]
&& \cup
\bigcup_{k=1}^{ \infty} \{(x,y) ~|~ x \in [\xi_k^+, \xi_{k-1}^+), y \in [0,1) \}.
\end{eqnarray}
See Fig.~{\ref{fig:NHBmap}}(I). Each piece of this partition 
$\{ (x,y) ~|~ x \in [\xi_k^{\pm}, \xi_{k-1}^{\pm}), y \in [0,1)\}$  
is uniformly stretched in the horizontal direction, and uniformly
squeezed in the vertical direction. The left pieces
$\{ (x,y) ~|~ x \in [\xi_k^{-}, \xi_{k-1}^{-}), y \in [0,1) \}$ 
are mapped to the bottom part of the unit cell and the right pieces
$\{ (x,y) ~|~ x \in [\xi_k^{+}, \xi_{k-1}^{+}), y \in [0,1) \}$ 
to the upper (Fig.~{\ref{fig:NHBmap}}(II)). 


Note that although our theoretical and numerical results are for the 
1-dimensional map $\phi(x)$, these results are also true for this
area-preserving map if an observable does not depend on the vertical
coordinate $y$, namely $A(x,y) = A(x)$, because the horizontal
coordinate $x$ of this area-preserving map is transformed by $\phi(x)$
and independent of $y$. Therefore this area-preserving extension of
$\phi(x)$ has also long time correlations with power law decay. This
fact contrasts to the results for the baker transformation, which
exhibits exponential decay of correlation functions.  

Here let us define some terms for later discussions. We define the
$k$-th escape domain ${\mathcal D}_{k}$ as
\begin{eqnarray}
{\mathcal D}_{k} 
= \{ (x,y) ~|~ x \in [\xi_{k-1}^{-}, \xi_{k-2}^{-}), y \in [0,1)\}
\end{eqnarray}
for $k=1,2,\cdots$. For the points in $k$-th escape domain 
${\mathcal D}_{k}$,  it takes $(k-1)$ times of the mappings $\psi^n$ to
escape from the left part $x<b$ to the right $x>b$. 
We also define the $k$-th injection domain 
${\mathcal D}_{k}^{\rm in}$ as  
\begin{eqnarray}
{\mathcal D}_{k}^{\rm in} 
= \{ \psi(x,y) ~|~ x \in [\xi_k^{+}, \xi_{k-1}^{+}), y \in [0,1) \}
\end{eqnarray}
for $k=1,2,\cdots$. The injection domains are the upper part of the
unit square, which are displayed in Fig.~{\ref{fig:NHBmap}}(II) by the
labels $\{A, B, C, \cdots\}$. We denote the areas, namely the Lebesgue
measures, of the $k$-th escape and injection domains as $S_k$ and 
$S_k^{\rm in}$, respectively. These areas obey scaling laws 
$S_k \sim 1/k^{\beta}$ and $S_k^{\rm in} \sim 1/k^{\beta+1}$. 
This latter power law gives the escape time distribution \cite{tm4}.

\section{\label{sec:summary}Summary and remarks}

In this paper, we have introduced a piecewise linear map and analyzed
its spectral properties. We have derived the generalized eigenfunctions
and eigenvalues explicitly for classes of observables and piecewise
constant initial densities. Our model is a modified version of the map
analyzed in Ref.~{\cite{tasaki1}}.  A main difference of these two
models is the normalizability of invariant densities.  The invariant
density of the model investigated in Ref.~{\cite{tasaki1}} is not
normalizable for a parameter region. This is a typical property of
dynamical systems with marginal fixed points 
\cite{tasaki1,aizawa,aaronson}, and is caused by divergence of invariant
density at marginal fixed points. 

On the other hand, the uniform density is invariant for the map
$\phi(x)$ discussed in the present paper; 
therefore the invariant density is normalizable for any values of the
system parameters, even though our system has also a marginal fixed
point. This is because the present model has the mechanism suppressing  
injections of the orbits into neighborhoods of the marginal fixed
point and this property prevents divergences of the invariant density  
at the marginal fixed point. As a consequence of the normalizability,
the present model does not exhibit non-stationarity, which is
generically observed in maps with marginal fixed points
\cite{tasaki1,aizawa,aaronson}. 

The spectral properties of the present model is similar to those of 
Ref.~{\cite{tasaki1}} in the locations of the discrete and the
continuous spectra. There are two simple eigenvalues $1$ and  
$\lambda_d \in (-1,~0)$; the former corresponds to the invariant
eigenstate and the latter to the oscillating one. The eigenstate
associated to $\lambda_d$, however, does not contribute to the long time
behaviors of the correlation functions because it decays exponentially
fast. There is also the continuous spectrum on the real interval
$[0,1]$; this continuous spectrum leads to power law decay of
correlation functions. We have confirmed a good  agreement between the 
theoretical prediction and the numerical result for scaling behaviors of  
the PSD $S(f) \sim 1/ f^{\gamma}$. 

Furthermore, the piecewise linear map $\phi(x)$ has been extended to an 
area-preserving invertible map on the unit square.  In contrast to
the baker  transformation, which is hyperbolic and shows exponential
decay of correlation functions, our model is non-hyperbolic and displays
power law decay of correlations. 

As is well known, the mixed type Hamiltonian systems often exhibit power
law decay of correlation functions. The area-preserving map $\psi(x,y)$
introduced in this paper may be considered as an abstract model of the
mixed type Hamiltonian systems in the following sense.  Instabilities of
the orbits of the map $\psi(x,y)$ [Eq.(\ref{eqn:2d-model})] is weak in
neighborhoods of the line $x=0$, and the escape time from the left part
$x<b$ to the right part $x > b$  diverges as $x \to 0$. In 
other words, the orbits stick to the line $x=0$ for long times. 
This property seems to be similar to dynamics of Hamiltonian
systems near torus, cantorus, and marginally unstable periodic orbits,
where chaotic orbits stick for long times. 

And, in fact, similar dynamics is observed in a Poincar\'e map of the
mushroom billiard \cite{tm4}, which has been proposed recently as a
model of mixed type systems with sharply divided phase spaces 
\cite{bunimovich,bunimovich2,altmann,altmann2,shudo}. In
Ref.~\cite{tm4}, it is found that an infinite partition can be
constructed on a Poincar\'e surface using escape times from
neighborhoods of the outermost tori, and that the area of the escape
domains and the injection domains obey the scaling relations 
${\mathcal D}_{k} \sim 1/k^2$ and 
${\mathcal D}_{k}^{\rm in} \sim 1/k^3$, respectively. These relations
correspond to the case $\beta = 2$ of the present model. Note that a
correlation function of the Poincar\'e map of the mushroom billiard
exhibits power law decay $C(n) \sim 1/n$ \cite{tm4}, and this is
consistent with the analytical result of the present paper. This
relation between the map $\psi(x,y)$ and a billiard system is similar to
that of the baker map and the Lorentz gas \cite{tel}. 

Since the map $\psi(x,y)$ is an elementary model of conservative systems
and can be treated analytically to some extent, this system may be
important for understanding relationships between non-equilibrium
phenomena, such as relaxation and transport, and underlying reversible
dynamics; this is a fundamental problem in dynamical system theory and
statistical mechanics 
\cite{gaspard2,dorfman}. 


\begin{acknowledgments}
 The authors would like to thank Prof.~S.~Tasaki for helpful suggestions 
 and discussions, and Prof.~A.~Shudo for valuable discussions and
 comments. This work is supported in part by Waseda University Grant for
 Special Research Projects (The Individual Research No.~2005B-243) from
 Waseda University. 
\end{acknowledgments}

\appendix*
\section{\label{sec:app}Asymptotic Behaviors}

\subsection{When \boldmath $|z| \to 0$}
In this appendix, we show the inequality,
\begin{eqnarray}
\left| \integ_0^{\infty} ds  
\f {s^{\beta -1} e^{-s}}{z - e^{-s}} \right|
< C
\left(
\log \f 1{|z|}
\right)^{\beta}
\label{app:lem}
\end{eqnarray}
as $|z| \to 0$ and ${\rm arg}z \in [0,2\pi]$ is fixed, where $C$ is a
positive constant. From this inequality, the Eqs.~(\ref{eqn:xito0}) and
(\ref{eqn:zto0}) can be derived.  Let $z = x+ iy$ in this subsection. 

First, let us assume $\arg z \in [3 \pi/4,~5\pi/4]$. Splitting the
integral into two pieces, we have
\begin{eqnarray} \nonumber
\hspace*{-.7cm}\integ_0^{\infty} ds 
\f {s^{\beta -1} e^{-s}}
   {\left|  z - e^{-s} \right|}  &
\\[.15cm]
&\hspace*{-1.2cm} \leq \integ_0^{\log \frac 1{|x|}} ds s^{\beta-1}
+\integ_{\log \frac 1{|x|}}^{\infty} ds 
 \f {s^{\beta -1} e^{-s}} {|x|}. 
\label{eqn:asymp1}
\end{eqnarray}
Apparently, the first term of the rhs of Eq.~(\ref{eqn:asymp1}) has an
upper bound $C_1\left( - \log  {|z|}\right)^{\beta}$ for some positive  
constant $C_1$.
On the other hand, the second term 
has an upper bound $C'_1\left( -\log {|z|}\right)^{\beta-1}$. This is
obtained by using an asymptotic expansion of the incomplete gamma
function (see e.g., Ref.~\cite{henrici}).  Thus, in this case the
inequality Eq.~(\ref{app:lem}) holds.

Second, we consider the case for $\arg z \in [\pi/4,~3 \pi /4]$ or 
$\arg z \in [5\pi/4,~7 \pi /4]$. Using similar calculations,
we have 
\begin{eqnarray} \nonumber
\integ_0^{\infty} ds 
 \f {s^{\beta -1} e^{-s}} 
    {\left| z - e^{-s} \right|} &\leq&
\left( \log \f 1{|y|} \right)^{\beta} 
\integ_0^{1} ds~
\f 1{\sqrt {H(s)} } \hspace*{1.5cm}\\[.3cm]
&&\hspace*{1.2cm}+
\integ_{\log \frac 1{|y|}}^{\infty} ds~
 \f {s^{\beta -1 }e^{-s}}{|y|},
\end{eqnarray}
 where we define $H(s)$ as
 $
 H(s) \equiv 
 \left( {x}/{|y|^s} - 1 \right)^2 + |y|^{2(1-s)}.
 $
 It is easy to check that $H(s) \geq 1/2$. Therefor, the first term has
 an upper bound  $C_2 \left( -\log {|z|} \right)^{\beta}$.
 By using the asymptotic expansion of the incomplete Gamma function, we
 have an upper bound for the second term:
 $C'_2 \left( -\log {|z|} \right)^{\beta-1}$.
 Thus, this case also satisfies the inequality Eq.~(\ref{app:lem}).

 Finally, we consider the case $\arg z \in [0,~\pi /4]$ or
 $\arg z \in [7\pi/4,~2\pi]$. We split the integral as 
\begin{widetext}
\begin{eqnarray}
\integ_0^{\infty} ds~ \f{s^{\beta-1} e^{-s}}{z - e^{-s}} =
 \integ_{0}^{x/2} dt 
\f {G(t)} {z - t}
+
\dip
\int_{0}^{x/2} dt
\left(
\f{ G(x-t) }{ t + i y} -
\f{ G(x+t) }{ t - i y} 
\right)
+ \integ_{3 x /2}^1 dt 
\f {G(t)} {z - t},
\label{app:case3-2}
\end{eqnarray}
\end{widetext}
where we define $G(t)$ as $G(t) \equiv \{\log (1/t)\}^{\beta-1}$.
 By using similar techniques as above, it can be shown that the 
 absolute values of the first and the third terms of the rhs of
 Eq.~(\ref{app:case3-2}) have an upper bound  
 $C_3 \left( -\log x\right)^{\beta}$.
 For the imaginary part of the second term, it is easy to derive
 an upper bound of its absolute value:
 $C'_3 \left( -\log x \right)^{\beta-1}$. 
 On the other hand, for the real part, we have an upper bound 
 $C'_3 \left( -\log x \right)^{\beta-2}$, which can be derived through
 integration by parts after changing the variables as $t' = t/x$. This
 completes the proof of Eq.~(\ref{app:lem}).  

\subsection{When \boldmath $|z| \to 1$}
 The property Eq.~(\ref{eqn:xito1}) is also derived in the same way. We
 briefly mention about the derivation. Let us begin with an analytic
 continuation of $\Xi (z)$
\begin{eqnarray}\nonumber
\Xi(z) &=& B_1^-(0) \Psi(z) \hspace*{5.55cm}\\[.15cm]
&& \hspace*{-.3cm} + \sum_{j=0}^{\infty} \sum_{l=2}^{\infty} 
 B_l^-(0) l^{\beta} \f {\rho_j}{\Gamma(\beta)} 
\int_0^{\infty} ds \f {s^{\beta -1} e^{-(j+l)s}}{z - e^{-s}}.
\label{eqn:app-Xi1}
\end{eqnarray}
We consider the second term. The absolute value of the second term has
 an upper bound
\begin{eqnarray}
{\rm const.} \times \left|
\sum_{j=0}^{\infty} \rho_j
\int_0^{\infty} ds \f {s^{\beta -1} e^{-(j+2)s}}{(z - e^{-s})(1 - e^{-s})}
\right|.
\label{eqn:app-Xi2}
\end{eqnarray}
Therefore, we show that for $k=1,2,\cdots$, and $1<\beta <2$,
\begin{eqnarray}
\left|
\int_0^{1} ds \f {s^{\beta -1} e^{-ks}}{(z - e^{-s})(1 - e^{-s})}
\right| 
< C' |1-z|^{\beta -2},
\label{app:lem2}
\end{eqnarray}
as $z \to 1$ and $\arg (z-1) \in [0, 2\pi)$ with a constant
$C'$. Note that the integral in Eq.~(\ref{eqn:app-Xi2}) from $1$ to
$\infty$ is convergent.  For $\beta \geq 2$, it can be analyzed
in a similar way but the rhs of Eq.~(\ref{app:lem2}) should be changed
to a constant ($\beta>2$) or a log correction $-\log |1-z|$
($\beta=2$). Let $z = 1 + x + iy$ in the following.  

First, let us assume $\arg z \in [\pi/4,~3 \pi /4]$ or 
$\arg z \in [5\pi/4,~7 \pi /4]$, then we have
\begin{eqnarray}
\int_0^{1} ds 
 \f {s^{\beta -1} e^{-ks}}
    {\left|
      (z - e^{-s})(1 - e^{-s})
     \right|
    }
 < 
e\int_0^1 ds 
 \f {s^{\beta -2}}
    {\left|
      z - e^{-s}
     \right|
    },  
\label{app2:case1}
\end{eqnarray}
where we have used $1-e^{-s} \geq s/e$ for $s \in [0,1]$. The rhs of 
Eq.~(\ref{app2:case1}) can be estimated, by splitting the integral, as  
\begin{eqnarray}\nonumber
\int_0^{1} ds \f {s^{\beta -2} }
 {\left|
   z - e^{-s} 
  \right|}
\leq \dip
\int_0^{2e|y|} ds 
\f {s^{\beta -2}}{ |y| } +
2e \int_{2e|y|}^{1} ds {s^{\beta -3} }, \\[-.35cm]
\label{app2:case1-2}
\end{eqnarray}
where we have used $1-e^{-s} \geq s/e$ again. It is obvious that the rhs
of Eq.~(\ref{app2:case1-2}) is less than $C'_4 |1-z|^{\beta-2}$. 
Therefore the inequality Eq.~(\ref{app:lem2}) is satisfied in this case.

Second, the case for  $\arg z \in [0,~\pi /4]$ or
$\arg z \in [7\pi/4,~2\pi)$ can be analyzed in the same way as the
first case (But, in this case, split the integral in terms of $x$
instead of $y$.). Thus we omit the detail. 

Finally, when $\arg z \in [3 \pi/4,~5\pi/4]$, we have
\begin{widetext}
\begin{eqnarray}
\integ_0^{1} ds~ \f{s^{\beta-1} e^{-ks}}{(z - e^{-s})(1 -e^{-s})} =
 \integ_{1/e}^{1-3\epsilon/2} dt 
\f {F(t)} {z - t}
+
\dip
\int_{0}^{\epsilon/2} dt
\left(
\f{ F(1-\epsilon-t) }{ t + i y} -
\f{ F(1-\epsilon+t) }{ t - i y} 
\right)
+ \integ_{1 - \epsilon /2}^1 dt 
\f {F(t)} {z - t},
\label{app2:case3-1}
\end{eqnarray}
\end{widetext}
where we define $\epsilon > 0$ as $\epsilon= |x|$ and $F(t)$ as 
$F(t) \equiv \{\log(1/t)\}^{\beta-1} t^{k-1}/(1-t)$. The first and the
third terms of the rhs can be estimated in the same way as
Eq.~(\ref{app2:case1-2}), and we obtain an upper bound for their absolute
values as $C_5 |1-z|^{\beta-2}$. For the imaginary part of the second
term, we have easily a bound of its absolute value:~
$C'_5 |1-z|^{\beta-2}$. On the other hand, 
for the real part, we also have a bound $C''_5 |1-z|^{\beta-2}$ through
integration by parts after changing the variables as 
$t' = t/ \epsilon$. Thus we complete the proof of Eq.~(\ref{app:lem2}).

From Eq.~(\ref{app:lem2}), it can be shown that 
$\Psi(z) \to \Psi(1) < \infty$, 
as $z \to 1$. Thus the first term of Eq.~(\ref{eqn:app-Xi1}) converges
as $z \to 1$. Consequently, we have Eq.~(\ref{eqn:xito1}). 
\vspace*{.2cm}
\bibliography{nhb}

\end{document}